\documentclass[12pt]{revtex4}
 
 \usepackage{hyperref}
 \usepackage{amsmath,amsfonts,amssymb}
 \hypersetup{dvips,dvipdfm,colorlinks=true,urlcolor=magenta,filecolor=magenta,linktoc=page,citecolor=red,linkcolor=blue,bookmarks=true}
 \usepackage{graphicx,epstopdf}
\begin{document}

\title{Non--minimally coupled scalar field in Kantowski--Sachs model and symmetry analysis}

\author{Sourav Dutta$^1$\footnote {sduttaju@gmail.com}}
\author{Muthusamy Lakshmanan $^2$\footnote {lakshman.cnld@gmail.com}}
\author{Subenoy Chakraborty$^3$\footnote {schakraborty.math@gmail.com}}
\affiliation{$^1$Department of Pure Mathematics, Ballygunge Science College, 35, Ballygunge Circular Rd, Ballygunge, Kolkata, West Bengal 700019\\
$^2$ Centre for Nonlinear Dynamics, Bharathidasan University, Tiruchirapalli - 620 024, India\\
$^3$Department of Mathematics, Jadavpur University, Kolkata-700032, West Bengal, India\\}


\begin{abstract}
The paper deals with a non--minimally coupled scalar field in the background of homogeneous but anisotropic Kantowski--Sachs space--time model. The form of the coupling function of the scalar field with gravity and the potential function of the scalar field are not assumed phenomenologically, rather they are evaluated by imposing Noether symmetry to the Lagrangian of the present physical system. The physical system gets considerable mathematical simplification by a suitable transformation of the augmented variables $(a, b, \phi)\rightarrow (u, v, w)$ and by the use of the conserved quantities due to the geometrical symmetry. Finally, cosmological solutions are evaluated and analyzed from the point of view of the present evolution of the Universe.	
\\\\

Keywords: Noether symmetry, Lie point symmetry, Non--minimally Coupled scalar field.

\end{abstract}
\maketitle

\section{introduction}
A scalar field non--minimally coupled to gravity is a very well known and widely used theory of induced gravity which is supposed to be a strong candidate in the context of unified theories \cite{mg}. The inflationary scenario in standard cosmology can be exhibited with this modified gravity theory without encountering any graceful exit problem \cite{ga}. Also the present gravity theory induces the Newtonian gravitational constant, the cosmological constant and overcomes the old problem of density perturbation \cite{rf}. Moreover, in quantum cosmology, the present induced gravity theory preserves the generic features of the wave functions due to  Hartle--Hawking \cite{jb} and Vilenkin \cite{av}. Astrophysically, it is possible to have worm hole solutions \cite{dh}, \cite{ak1} in this induced gravity theory. Although there are a lot of successful applications of this theory, but still one cannot determine the coupling function and the potential function analytically from any physical principle and hence they are usually chosen in an adhoc basis.\\

On the otherhand, it may be possible to have analytic expressions for both the coupling function and the potential function by imposing geometric (i.e., Lie point/ Noether) symmetries. Mathematically, Lie point/ Noether symmetries can be used to examine the self consistency of phenomenological physical models in one hand and also one may be able to restrict physical parameters (namely equations of state parameter) or arbitrary functions (potential of a scalar field). Further, if the Lagrangian of the system possesses Noether symmetries then the physical system is associated with some conserved quantities and as a result, the evolution equations are simplified to a great extent or even they may be solvable.\\

The present work is an attempt of using Lie point and Noether symmetries to the non--linear coupled modified Einstein field equations for non--minimally coupled scalar field cosmology. Basically, we want to proceed with the problem from two different angles: (i) analytic evaluation of the unknown functions of the scalar field by the imposition of symmetry and (ii) by a suitable change of augmented variables to have at least one cyclic co--ordinate of the Lagrangian dynamics, so that the dynamical equations can be solved and may have interesting cosmological implications. It should be mentioned that this problem was studied in reference \cite{ak} and it was found that due to Noether symmetry the Lagrangian of the system become degenerate. In this work, we shall show that by introducing the Lapse function in the augmented space it is possible to have Noether symmetry and the Lagrangian will no longer be degenerate. 

The paper is organized as follows: Section II deals with the Lie point symmetries of the induced gravity theory. Noether symmetry is imposed in section III and cosmological solutions are determined after making canonical transformation of the dynamical variables. At the end cosmological implications and summary of the work is presented in section IV.

\section {Lie point symmetries of the induced gravity theory}

The action integral for the scalar field, non--minimally coupled to gravity and having potential function $V(\phi),$ is given by 

\begin{equation}
A=\int d^4 x \sqrt{-g}\left[f(\phi)R-\frac{1}{2}\phi_{,\mu}\phi^{,\mu}-V(\phi)\right],\label{k1}
\end{equation}

where $R$ is the usual Ricci scalar and $g$ is the determinant of the metric tensor. Here the coupling function $f(\phi)$ and the potential $V(\phi)$ will be determined by imposing geometric symmetries to the system. In the background of homogeneous but anisotropic Kantowski--Sachs space--time model, i.e,
\begin{equation}
ds^2=-N^2 dt^2+a^2(t) dr^2+b^2(t) d\Omega_2^2,\label{k2}
\end{equation}

 the above  action has the following explicit expressions (neglecting the surface term)
 
 
 \begin{equation}
 A=4\pi\int \left[\frac{-4f'ab\dot{b}\dot{\phi}}{N}-\frac{2f'b^2\dot{a}\dot{\phi}}{N}-\frac{4fb\dot{a}\dot{b}}{N}-\frac{2fa\dot{b}^2}{N}+2Nfa+\frac{1}{2N}ab^2\dot{\phi}^2-Nab^2V(\phi)\right]dt,\label{k3}
 \end{equation} 
 where $N$ is known as the Lapse function and $a$ and $b$ are the metric coefficients in the radial and angular directions.\\

Now varying the action with respect to the augmented variables $(a, b, \phi)$ we have the evolution equations:

\begin{equation}
\frac{\ddot{a}}{a}+\frac{\ddot{b}}{b}+\frac{f'}{f}\ddot{\phi}+\frac{\dot{a}}{a}\frac{\dot{b}}{b}+\frac{f'}{f}\frac{\dot{a}}{a}\dot{\phi}+\frac{f'}{f}\frac{\dot{b}}{b}\dot{\phi}+\frac{f''}{f}\dot{\phi}^2+\frac{\dot{\phi}^2}{4f}-\frac{V(\phi) N^2}{2f}=0,\label{k4}
\end{equation}

\begin{equation}
\frac{\ddot{a}}{a}+2\frac{\ddot{b}}{b}+2\frac{\dot{a}}{a}\frac{\dot{b}}{b}+\frac{\dot{b}^2}{b^2}-\frac{\ddot{\phi}}{2f'}-\left(\frac{\dot{a}}{a}+2\frac{\dot{b}}{b}\right)\frac{\dot{\phi}}{2f'}+\frac{N^2}{b^2}-\frac{V'(\phi) N^2}{2f'}=0,\label{k5}
\end{equation}

\begin{equation}
2\frac{\ddot{b}}{b}+\frac{f'}{f}\ddot{\phi}+\frac{f''}{f}\dot{\phi}^2+2\frac{f'}{f}\frac{\dot{b}}{b}\dot{\phi}+\frac{\dot{b}^2}{b^2}+\frac{\dot{\phi}^2}{4f}+\frac{N^2}{b^2}
-\frac{V(\phi)N^2}{2f}=0,\label{k6}
\end{equation}

  and varying the action with respect to the Lapse function $N$ we have the constraint equation relating the above evolution equations as:

\begin{equation}
\frac{\dot{b}^2}{b^2}+\frac{f'}{f}\frac{\dot{a}}{a}\dot{\phi}+2\frac{f'}{f}\frac{\dot{b}}{b}\dot{\phi}-\frac{\dot{\phi}^2}{4f}+2\frac{\dot{a}}{a}\frac{\dot{b}}{b}+\frac{N^2}{b^2}-\frac{V(\phi)N^2}{2f}=0,\label{k7}
\end{equation}

where an overhead prime indicates differentiation with respect to the scalar field $\phi$ while an over dot implies time differentiation of the variable. We shall now impose the existence of Lie point symmetries to the present problem. The basic theory of this (Lie) point symmetry is as follows:\\

For a system of second order ordinary differential equations of the form 

\begin{equation}
\ddot{x}^i=w^i(t, x^j, \dot{x}^j), ~i=j=1,2,3,......,n\label{k8}
\end{equation}
 
a vector field $X=\xi \frac{\partial}{\partial t}+\eta^i \frac{\partial}{\partial x^i}$ ($\xi, \eta^i$ are functions of $t$ and $x^i$ i.e., $\xi=\xi(t, x^i)$, $\eta^i=\eta^i(t, x^i))$ is termed as the generator of a Lie point symmetry provided \cite{gb}

\begin{equation}
X^{[2]}\big(\ddot{x}^i-w^i(t, x^j, \ddot{x}^j)\big)=0,\label{k9}
\end{equation}

where \begin{equation}
X^{[2]}=\xi \partial_t+\eta^i \partial_i+\big(\dot{\eta}^i-\dot{x}^i \dot{\xi}\big)\partial \dot{x}^i+\left(\ddot{\eta}^i-\dot{x}^i\ddot{\xi}-2\ddot{x}^i\dot{\xi}\right)\partial \ddot{x}^i,\label{k10}
\end{equation}
 is known as the second prolongation of $X$ (with $\partial_u \equiv \frac{\partial}{\partial u}$). One can equivalently state the Lie point symmetry condition as

\begin{equation}
\big[X^{[1]}, A\big]=\lambda(x^i) A,\label{k11}
\end{equation}

where 
\begin{equation}
X^{[1]}=\xi \partial_t+\eta^i \partial_i+\big(\dot{\eta}^i-\dot{x}^i \dot{\xi}\big)\partial \dot{x}^i,\label{k12}
\end{equation}

and
\begin{equation}
A=\partial_t+\dot{x}^i \partial_{x^{i}}+w^i\big(t, x^j, \dot{x}^j\big)\partial_{\dot{x}^j},\label{k13}
\end{equation}

are respectively termed as the first prolongation vector of $X$ and the Hamiltonian vector field. It should be noted that in symmetry analysis the combined space of both independent and dependent variables is termed as augmented space. So in the present problem the augmented space $(t, a, b, \phi)$ is a 4 dimensional space. Then as an extension of the augmented space one can define the jet space where the co-ordinates consist of not only the dependent and independent variables but also the partial derivatives of the dependent variables. In the present problem, as we are dealing with second order ordinary differential equations, the jet space contains partial derivatives up to second order. As a consequence the infinitesimal generator $X$ has been extended as the prolongated vector field $X^{[2]}$ over the jet space \cite{sd1}. Now, the evolution equations (\ref{k4})--(\ref{k6}) can be recast in the form (\ref{k8}) as follows:

\begin{eqnarray}
\ddot{a}&\equiv&w_1\big(t, a, \dot{a}, b, \dot{b}, \phi, \dot{\phi} \big)= a\left[f_2-f_3+\frac{2f'^2-f}{6 f'^2+2f}\left(f_3-2(f_2-f_1) \right) \right]\nonumber\\
\ddot{b}&\equiv& w_2\big(t, a, \dot{a}, b, \dot{b}, \phi, \dot{\phi} \big)=\frac{b}{2}\left[f_3-\frac{f'^2}{3 f'^2+f}\left(f_3-2(f_2-f_1) \right) \right]\nonumber\\
\ddot{\phi}&\equiv& w_3\big(t, a, \dot{a}, b, \dot{b}, \phi, \dot{\phi} \big)=\frac{ff'}{3 f'^2+f}\left(f_3-2(f_2-f_1) \right),\label{k14}
\end{eqnarray}
where $$f_1=\frac{\dot{a}}{a}\frac{\dot{b}}{b}+\frac{f'}{f}\frac{\dot{b}}{b}\dot{\phi}-\frac{f''}{f}\dot{\phi}^2-\frac{\dot{\phi}^2}{2f}+\frac{N^2}{b^2}+\frac{\dot{b}^2}{2f},$$
$$f_2=\left(\frac{\dot{a}}{a}+2\frac{\dot{b}}{b}\right)\frac{\dot{\phi}}{2f'}-\frac{V(\phi)N^2}{2f}+\frac{V'(\phi)N^2}{2f'}+\frac{f'}{f}\frac{\dot{a}}{a}\dot{\phi}+2\frac{f'}{f}\frac{\dot{b}}{b}\dot{\phi}-\frac{\dot{\phi}^2}{4f},$$
and
$$f_3=-\frac{\dot{\phi}^2}{2f}-\frac{f''}{f}\dot{\phi}^2+\frac{f'}{f}\frac{\dot{a}}{a}\dot{\phi}+2\frac{\dot{a}}{a}\frac{\dot{b}}{b}.$$
Hence the infinitesimal generator of the point transformation for the set of equation (\ref{k14}) can be written as

\begin{equation}
	X=\xi \frac{\partial}{\partial t}+\eta_1 \frac{\partial}{\partial a}+\eta_2 \frac{\partial}{\partial b}+\eta_3 \frac{\partial}{\partial \phi}+\eta'_1 \frac{\partial}{\partial \dot{a}}+\eta'_2 \frac{\partial}{\partial \dot{b}}+\eta'_3 \frac{\partial}{\partial \dot{\phi}},\label{k15}
\end{equation}

where $\xi=\xi(t, a, b, \phi),~\eta_i=\eta_i(t, a, b, \phi),~i=1, 2, 3.$ Also using the total derivative operator as 

\begin{equation}
	\frac{d}{dt}=\frac{\partial}{\partial t}+\dot{a}\frac{\partial}{\partial a}+\dot{b}\frac{\partial}{\partial b}+\dot{\phi}\frac{\partial}{\partial \phi},\label{k16}
\end{equation}

we write

\begin{equation}
	\eta'_i=\frac{d \eta_i}{dt}-\alpha_i\frac{d \xi}{dt},~~~~\alpha_i=(a, b, \phi),~ i=1, 2, 3.\label{k17}
\end{equation}

Thus the Lie point symmetry conditions corresponding to the set of equation (\ref{k14}) can be written as \cite{gb}

\begin{equation}
Xw_i=\eta''_i,~~i=1, 2, 3,\label{k18}
\end{equation}

with 

\begin{equation}
\eta_i''=\frac{d \eta_i''}{dt}-\ddot{\alpha_i}\frac{d \xi}{dt}=\frac{d \eta_i''}{dt}-w_i\frac{d \xi}{dt}.\label{k19}
\end{equation}

Solving the equation (\ref{k14}) corresponding to the Lie symmetry conditions (\ref{k18}) we have the following solutions of the overdetermined system of equations,

\begin{equation}
\xi=\xi_0(t+1),~\eta_1=\alpha a,~\eta_2=\beta b,~\eta_3=-\gamma \phi,\label{k20}
\end{equation} 
with $f=f_0 \phi^2$ and $V=V_0 \phi^4$, where $\xi_0, \alpha, \beta, \gamma, f_0$ and $V_0$ are all constants.\\
Note that the symmetry solution (\ref{k20}) is valid for $f_0\neq -\frac{1}{12}.$ For $f_0=-\frac{1}{12}$ the Hessian of the Lagrangian vanishes identically and the present system reduces to a singular system.
 So for $f_0\neq -\frac{1}{12}$ the symmetry vector is
\begin{equation}
X=\xi_0(1+t)\partial_t+\alpha a \partial_a+\beta b \partial_b-\gamma \phi \partial_{\phi}, \label{k21}
\end{equation}
with $\xi_0, \alpha, \beta, \gamma, f_0$ and $V_0$ being arbitrary constants.\\

Hence the characteristic equation corresponding to this symmetry can be written as 
\begin{equation}
\frac{dt}{\xi_0(1+t)}=\frac{da}{\alpha a}=\frac{db}{\beta b}=\frac{d\phi}{-\gamma \phi}. \label{k22}
\end{equation}
Thus one has the parametric solution
\begin{equation}
a=a_0(t+l),~b=b_0(t+m),~\phi=\frac{\phi_0}{(t+n)},\label{k23}
\end{equation}
where $a_0, b_0, \phi_0$ are integration constants and
\begin{equation}
l=\frac{\alpha}{\xi_0},~m=\frac{\beta}{\xi_0},~n=\frac{\gamma}{\xi_0}.\label{k24}
\end{equation}
Using the above parametric solution into the field equations (\ref{k4})--(\ref{k7}) we have
\begin{equation}
l=2,~m=n=1,~\frac{N^2}{b_0^2}=1,~V_0\phi_0^2N^2=8(1+f_0).\label{k25}
\end{equation}
This parametric family of solution corresponding to the above Lie symmetry gives simple power law cosmological solution which is not of much physical interest in the present context.

\section{The Noether symmetry approach and Cosmological Solutions}

The Noether symmetry corresponds to the invariance of the Lagrangian of the physical system under the Lie derivative along an appropriate vector field. Also by imposing this symmetry, one gets a conserved quantity associated with the system and, as a result, it is possible to simplify the equations of motion or even solve them (\cite{scap1}, \cite{scap2}, \cite{scap3}, \cite{scap4},  and for review \cite{scap5} ).\\

In the general framework for a point like canonical Lagrangian , $L\big(q^{\alpha}(x^i), \partial_j q^{\alpha}(x^i)\big)$, one can contract the Euler--Lagrange equations

\begin{equation}
\partial_j\big(\frac{\partial L}{\partial \partial_j q^{\alpha}}\big)=\frac{\partial L}{\partial q^{\alpha}},\label{k26}
\end{equation}

with some unknown functions, $\mu^\alpha(q^\beta,)$ i.e,

\begin{equation}
\mu^\alpha\left[\partial_j\left(\frac{\partial L}{\partial \partial_j q^\alpha}\right)-\frac{\partial L}{\partial q^\alpha}\right]=0,\label{k27}
\end{equation}

and a simplification gives 

\begin{equation}
\mathcal{L}_{\overrightarrow{X}}L=\left[\mu^\alpha\frac{\partial }{\partial q^\alpha}+(\partial_j \mu^{\alpha})\frac{\partial }{\partial \partial_j q^\alpha}\right]L=\partial_j\left(\mu^\alpha\frac{\partial L}{\partial \partial_j q^\alpha}\right).\label{k28}
\end{equation}

Here notationally the Lie derivative with respect to the vector field

\begin{equation}
\overrightarrow{X}=\mu^\alpha\frac{\partial}{\partial q^\alpha}+\left(\partial_j \mu^\alpha\right)\frac{\partial}{\partial \partial_j q^\alpha},\label{k29}
\end{equation}

is denoted by $\mathcal{L}_{\overrightarrow{X}}$.\\

According to Noether, the vector field $\overrightarrow{X}$ can be chosen in such a way that the Lagrangian of the system is invariant along the vector field, i.e, $\mathcal{L}_{\overrightarrow{X}}L=0.$ Then we say that the physical system is invariant under Noether symmetry with $\overrightarrow{X}$ as the infinitesimal generator of the symmetry. Further, we see that due to Noether symmetry, the physical system has a constant of motion namely \cite{sr6} 

\begin{equation}
Q^j=\mu^\alpha\frac{\partial L}{\partial \partial_j q^\alpha},\label{k30}
\end{equation}

which is conserved as 

\begin{equation}
\partial_i Q^i=0.\label{k30a}
\end{equation}

Here $Q^i$ is also termed as Noether charge( or conserved charge). Also the energy function associated with the Lagrangian is 

\begin{equation}
E=\frac{\partial L}{\partial {q}^\alpha}\dot{q}^\alpha-L,\label{k31}
\end{equation}

i.e., $E$ is the total energy $(T+V)$ of the system and is a constant of motion \cite{sr2}.\\

We shall now show how by imposing the existence of Noether symmetry the unknown coupling function and the potential of the non--minimally coupled scalar field can be determined. Further, by suitable transformation of the augmented variables $(a, b, \phi)$ it is possible to have at least one cyclic co--ordinate so that the evolution equations are reduced and the dynamics can be solved exactly. Moreover, one can use the Noether symmetry approach to select reliable models \cite{sr4} provided the associated conserved quantities have some physical meaning.\\

 The Lagrangian for the present non--minimally coupled scalar field in Kantowski--Sachs--Space--time is given by (see equation (\ref{k3}))
 
 \begin{equation}
 L=-\frac{4f'ab \dot{b}\dot{\phi}}{N}-\frac{2f'b^2\dot{a} \dot{\phi}}{N}-\frac{4fb\dot{a} \dot{b}}{N}-\frac{2fa \dot{b}^2}{N}+2Nfa+\frac{1}{2N}ab^2\dot{\phi}^2-ab^2V(\phi)N,\label{k32}
 \end{equation}
 
 and the corresponding infinitesimal generator of the Noether symmetry has the explicit form \cite{mt3}

\begin{equation}
\overrightarrow{X}=\alpha\frac{\partial}{\partial a}+\beta\frac{\partial}{\partial b}+\gamma\frac{\partial}{\partial \phi}+\dot{\alpha}\frac{\partial}{\partial \dot{a}}+\dot{\beta}\frac{\partial}{\partial \dot{b}}+\dot{\gamma}\frac{\partial}{\partial \dot{\phi}}+\delta\frac{\partial}{\partial N},\label{k33}
\end{equation}

where $\alpha, \beta, \gamma$ and $\delta$ are functions in the augmented space, i.e, $\alpha=\alpha(a, b, \phi)$ and so on while $\dot{\alpha}=\frac{\partial \alpha}{\partial a}\dot{a}+\frac{\partial \alpha}{\partial b}\dot{b}+\frac{\partial \alpha}{\partial \phi}\dot{\phi}$ and similarly for $\dot{\beta}$ and $\dot{\gamma}$.\\

Now by imposing the Noether symmetry condition $\mathcal{L}_{\overrightarrow{X}}L=0$ we have a set of first order partial differential equations:

\begin{equation}
2f \frac{\partial \beta}{\partial a} +f'b \frac{\partial \gamma}{\partial a}=0,\label{k34}
\end{equation} 

\begin{equation}
\alpha+a \gamma \frac{f'}{f}+2b \frac{\partial \alpha}{\partial b}+2a \frac{\partial \beta}{\partial b}+2ab \frac{\partial \gamma}{\partial b}-\frac{\delta}{N}a=0,\label{k35}
\end{equation}

\begin{equation}
\frac{\alpha b}{2}+\beta a-2f'b \frac{\partial \alpha}{\partial \phi}-4f'a\frac{\partial \beta}{\partial \phi}+ab \frac{\partial \gamma}{\partial \phi}-\frac{\delta ab}{2N}=0,\label{k36}
\end{equation}

\begin{equation}
\beta+b \frac{\partial \alpha}{\partial a}+a \frac{\partial \beta}{\partial a}+b \frac{\partial \beta}{\partial b}+b \gamma \frac{f'}{f}+ab\frac{\partial \gamma}{\partial a}\frac{f'}{f}+\frac{1}{2}b^2\frac{\partial \gamma}{\partial b}\frac{f'}{f}-\frac{\delta}{N}b=0,\label{k37}
\end{equation}

\begin{equation}
f\left(b \frac{\partial \alpha}{\partial \phi}+a \frac{\partial \beta}{\partial \phi}\right)+f'\left(b \alpha+a \beta+\frac{b^2}{2} \frac{\partial \alpha}{\partial b}+ab \frac{\partial \beta}{\partial b}+ab \frac{\partial \gamma}{\partial \phi}\right)+f''ab \gamma-\frac{ab^2}{4} \frac{\partial \gamma}{\partial b}-\frac{\delta}{N}abf'(\phi)=0,\label{k38}
\end{equation}

\begin{equation}
f\frac{\partial \beta}{\partial \phi}+f'\left(\beta+\frac{b}{2} \frac{\partial \alpha}{\partial a}+a\frac{\partial \beta}{\partial a}+\frac{b}{2} \frac{\partial \gamma}{\partial \phi}\right)+\frac{f'' b \gamma}{2}-\frac{ab}{4} \frac{\partial \gamma}{\partial a}-\frac{\delta}{2N}f' b=0,\label{k39}
\end{equation}

and the potential determining equation is

\begin{equation}
f \alpha+f' a \gamma-\frac{ab^2}{2}\left[V\left(\frac{\alpha}{a}+2\frac{\beta}{b}\right)+V'\gamma\right]+\frac{\delta a}{N}\left(f-\frac{1}{2}b^2V(\phi)\right)=0.\label{k40}
\end{equation}

Using the standard technique of separation of variables one can solve the above coupled partial differential equations to obtain the coefficients of the infinitesimal generator and the unknown functions of the theory as 

\begin{equation}
\alpha=\alpha_0 a,~ \beta=\alpha_0 b,~\gamma=-\alpha_0 \phi,~\delta=\alpha_0 N,~\mbox{ and } f=f_0\phi^2,~V=V_0\phi^4,\label{k41}
  \end{equation}

with $\alpha_0, f_0, V_0$ as constants of integration.\\

It should be noted that in deriving the Lie symmetry we have assumed the lapse function $N=1$ for simplicity so that in the infinitesimal generator (\ref{k21}) there is no partial derivative $\frac{\partial}{\partial N}$. On the other hand, in evaluating infinitesimal generator for Noether symmetry, we have considered the partial derivative $\frac{\partial}{\partial N}$ as non--zero as the constraint equation (\ref{k7}) is obtained from the Lagrangian (\ref{k32}) by considering variation with respect to $N$. In any case putting $\delta=0$ in the infinitesimal generator for the Noether symmetry we see that the Noether symmetry is a particular case of the Lie symmetry (By choosing $\beta=\alpha_0$ and $\gamma=-\alpha_0$). Hence we can say that the Lie algebra of the Noether symmetry is a subalgebra of the Lie algebra of the Lie symmetry obtained in the previous section. It should be noted that in reference \cite{ak} it is claimed that the Noether point symmetry is not possible due to degenerate Lagrangian. Here we have shown that by considering the lapse function as one of the augmented variables (we have used the constraint equation (\ref{k7})) properly and as a result the degeneracy of the Lagrangian is removed and we have obtained infinitesimal symmetry generators both for the Lie and Noether symmetries. Also the conserved charge (defined in equation (\ref{k30})) and conserved energy ({\it i.e.,} equation (\ref{k31})) have the explicit form 

\begin{equation}
Q=\alpha \frac{\partial L}{\partial \dot{a}}+\beta \frac{\partial L}{\partial \dot{b}}+\gamma \frac{\partial L}{\partial \dot{\phi}},\label{k42}
\end{equation}

and

\begin{equation}
E=\frac{ab^2\phi^2}{N}\left[\frac{1}{6}\frac{\dot{b}^2}{b^2}+\frac{1}{3}\frac{\dot{a}}{a}\frac{\dot{b}}{b}+\frac{1}{2}\left(\frac{\dot{\phi}}{\phi}\right)^2+N^2V_0\phi^2+\frac{N^2}{6}b^{-2}+\frac{2}{3}\frac{\dot{b}}{b}\frac{\dot{\phi}}{\phi}+\frac{1}{3}\frac{\dot{a}}{a}\frac{\dot{\phi}}{\phi}\right].\label{k43}
\end{equation}

In the next part of this section we shall determine a point transformation in the augmented space so that one of the transformed dynamical variables becomes cyclic and as a result the evolution equations from the transformed Lagrangian are easier to solve.\\

At first, we introduce the cartan one form \cite{sr2} namely 

\begin{equation}
\theta_L=\frac{\partial L}{\partial \dot{a}}da+\frac{\partial L}{\partial \dot{b}}db+\frac{\partial L}{\partial \dot{\phi}}d\phi,\label{k44}
\end{equation}

so that the conserved charge can be written as the contraction between the vector field $\overrightarrow{X}$  and the differential form $\theta_L$
 
\begin{equation}
Q=i_{\overrightarrow{X}} \theta_L.\label{k45}
\end{equation}

Now due to the point transformation :$(a, b, \phi, N)\rightarrow (u, v, w, \tilde{N})$ the infinitesimal generator $X$ gets transformed to

\begin{eqnarray}
\overrightarrow{\widetilde{X}}&=&(i_{\overrightarrow{X}} du)\frac{\partial}{\partial u}+(i_{\overrightarrow{X}}dv)\frac{\partial}{\partial v}+(i_{\overrightarrow{X}} dw)\frac{\partial}{\partial w}+(i_{\overrightarrow{X}} dN)\frac{\partial}{\partial N}\nonumber\\
&+&\left(\frac{d}{dt}(i_{\overrightarrow{X}}du)\right)\frac{d}{d \dot{u}}\nonumber
+\left(\frac{d}{dt}(i_{\overrightarrow{X}}dv)\right)\frac{d}{d\dot{v}}+\left(\frac{d}{dt}(i_{\overrightarrow{X}}dw)\right)\frac{d}{d \dot{w}},\label{k46}
\end{eqnarray}

so that $\overrightarrow{\widetilde{X}}$ can be considered as the lift of a vector field defined on the augmented space.\\

Further, as $\overrightarrow{X}$ is a symmetry vector so we can restrict the above point transformation such that 

\begin{equation}
i_{\overrightarrow{X}} du=1~~\mbox{and}~~ i_{\overrightarrow{X}}dv=0=i_{\overrightarrow{X}}dw=i_{\overrightarrow{X}}dN.\label{k47}
\end{equation}

As a result $\overrightarrow{\widetilde{X}}$ takes the simplified form $\overrightarrow{\widetilde{X}}=\frac{\partial}{\partial u}$ and we have $\frac{\partial}{\partial u}=0,$ i.e, $u$ is a cyclic co--ordinate. Hence the dynamics of the system can be reduced \cite{sr5}. The solution of the resulting first order linear partial differential equations gives the explicit transformations,

\begin{equation}
e^u=ab\phi,~e^v=a\phi,~e^w=b\phi,~e^{\tilde{N}}=\frac{ab^2\phi^2}{N},\label{k48}
\end{equation}

{\it i.e.,}
\begin{equation}
a=e^{u-w},~b=e^{u-v},~\phi=e^{v+w-u}.\label{k49}
\end{equation}

The conserved quantities in terms of the new variables are given by 

\begin{equation}
Q=\frac{\alpha_0}{2}\left(\dot{v}+\dot{w}-\dot{u}\right),\label{k50}
\end{equation}

and

\begin{equation}
E=\frac{1}{2} \left[2\dot{u}\dot{w}+2\dot{u}\dot{v}-\dot{u}^2-\dot{v}^2+e^{2(v+w)}+4V_0e^{2v+4w}\right].\label{k51}
\end{equation}

The transformed Lagrangian takes the form

\begin{equation}
L=\frac{1}{2} \left[2\dot{u}\dot{w}+2\dot{u}\dot{v}-\dot{u}^2-\dot{v}^2-e^{2(v+w)}-4V_0e^{2v+4w}\right],\label{k52}
\end{equation}
where we have chosen $f_0=-\frac{1}{8}$ and $e^{\xi}=1$ for simplicity.\\

Now using the conserved quantities the evolution equations from the above transformed Lagrangian can be solved easily as(for details see the appendix)

\begin{eqnarray}
a&=&\sqrt{t^2+v_0^2}e^{\left[\left(\lambda_0-2\frac{Q_0}{\alpha_0}\right)t+\lambda_1-\frac{t}{v_0}\tan^{-1}\left(\frac{t}{v_0}\right)\right]}\nonumber\\
b&=&te^{-\frac{2Q_0}{\alpha_0}t}\nonumber\\
\phi&=&e^{\frac{2Q_0}{\alpha_0}t},\label{k53}
\end{eqnarray}
with $Q_0, \alpha_0, v_0$ as integration constants.\\



In the following section we shall discuss the above solution from the point of view of cosmic evolution.

\section{Cosmological implications and Summary} 

The above solution (in equation (\ref{k53})) describe three types of cosmic evolution depending on the sign of $Q_0$ and $\mu_0=\lambda_0-\frac{2Q_0}{\alpha_0}$, and are named as type-I ($Q_0<0, \mu_0>0 $), type-II ($Q_0>0, \mu_0<0 $) and type-III ($Q_0>0, \mu_0>0 $) solutions respectively. There are two types of bouncing cosmic scenario depending on the sign of $\mu_0$. If $\mu_0>0$ (for type-I and type-III) the Universe expands at the beginning, reaches a maximum volume (bouncing point) and finally it contracts to big crunch. Type-I solution starts from a decelerating phase and then changes its sign several times and finally the Universe will be in an acceleration era. For type-III solution the Universe was initially in an accelerated expansion, then there is a decelerating phase and finally the Universe will be in an accelerating phase. The cosmic scenario is opposite when  $\mu_0<0$. In this case (type-II solution) the Universe initially contracts, reaches a minimum volume (bouncing point) and then expands exponentially. However, the Universe is always in an accelerating phase. Hence the Type-II cosmological solution is not physically realistic. \\

Moreover, as the present cosmic solutions represent singularity free bouncing cosmological scenario so they do not match with the big bang standard cosmology. However, type-III solution is rather interesting. Here the Universe starts from an inflationary era, then there is a phase of decelerated evolution and finally the Universe undergoes an accelerated evolution as predicted by recent observations \cite{Ag}, \cite{sp}, \cite{ec}, \cite{dj}, \cite{par} . Therefore the present work is an example where the evolution equations are highly coupled non--linear differential equations, very hard to solve with usual techniques but solutions are obtained by imposing symmetries (Lie and Noether) to them. Further, unknown functions in the physical theory are not estimated phenomenologically, rather they are also determined from the symmetry conditions. Finally, we have shown that the claim in reference \cite{ak} regarding degeneracy of the Lagrangian can be overcome by including the Lapse variable in the augmented space.

\section*{Acknowledgments} 
Author SD acknowledge Science and Engineering Research Board (SERB),
Govt. of India, for awarding National Post-Doctoral Fellowship
(File No: PDF/2016/001435) and the Department
of Mathematics, Jadavpur University where a part
of the work was completed.  Author ML thanks the National Academy of Sciences, India for the award of a Platinum Jubilee Senior Scientist Fellowship.. SC thanks Inter University Center for Astronomy and Astrophysics (IUCAA), Pune, India for their
warm hospitality as a part of the work was done during a visit. Also SC thanks UGC-DRS programme at the Department of Mathematics, Jadavpur University.

 \section*{Appendix}
The Lagrangian in the new variables $(u, v, w)$ takes the form

\begin{eqnarray}
L=\frac{1}{2} \left[2\dot{u}\dot{w}+2\dot{u}\dot{v}-\dot{u}^2-\dot{v}^2-e^{2(v+w)}-4V_0e^{2v+4w}\right].\label{a1}
\end{eqnarray}
Note that $u$ is the cyclic co--ordianate. So\\
\begin{eqnarray}
\frac{\partial L}{\partial \dot{u}}&=&\mbox{constant}.\nonumber\\
{\it i.e.,}~\dot{v}+\dot{w}-\dot{u}&=&\mbox{constant}=c (\mbox{say}).\nonumber\\
{\it i.e.,}~Q&=&\mbox{constant},\label{a2}
\end{eqnarray}
which nothing but the conserved Noether charge. The other two Euler--Lagrange equations are
\begin{eqnarray}
\ddot{u}-\ddot{v}+e^{2(v+w)}+4V_0 e^{2v+4w}=0,\label{a3}
\end{eqnarray}
and
\begin{eqnarray}
\ddot{u}+e^{2(v+w)}+8V_0 e^{2v+4w}=0,\label{a4}
\end{eqnarray}
Using (\ref{a2}) in (\ref{a3}) and (\ref{a4})) to eliminate $\ddot{u}$ we obtain
\begin{equation}
\ddot{v}+4V_0 e^{2v+4w}=0,\label{a5}
\end{equation}
\begin{equation}
\ddot{w}+e^{2(v+w)}+4V_0 e^{2v+4w}=0,\label{a6}
\end{equation} 
$${\it i.e.,}~ \frac{\ddot{v}}{e^{2v}}=-4V_0e^{4w},$$
and
$$\frac{\ddot{w}}{e^{2v}}=-e^{2w}-4V_0e^{4w},$$
$${\it i.e.,}~ \frac{\ddot{w}}{\ddot{v}}=\frac{1+4V_0e^{2w}}{4V_0e^{2w}},$$
$${\it i.e.,}~ \frac{\ddot{w} 4V_0e^{2w}}{1+4V_0e^{2w}}=\ddot{v}=f(t) (\mbox{say})$$
where $f(t)$ is an arbitrary function of ``$t$. "\\
To obtain a solution, we choose $e^w=w_0 t^n$, then the arbitrary function $f(t)$ becomes
\begin{equation}
f(t)=-\frac{n t^{2n-2}}{t^{2n}+\frac{1}{4V_0w_0^2}},\label{a7}
\end{equation} 
so that the differential equation for $v$ becomes 
\begin{equation}
\ddot{v}=-\frac{n t^{2n-2}}{t^{2n}+\frac{1}{4V_0w_0^2}}.\label{a8}
\end{equation} 
In particular, for $n=1$
$$\ddot{v}=-\frac{1}{t^{2}+\frac{1}{4V_0w_0^2}},$$
which has a first integral
$$\dot{v}=-\frac{1}{v_0} \tan^{-1}\left(\frac{t}{v_0}\right)+\lambda_0,~v_0^2=\frac{1}{4V_0w_0^2}.$$
Integrating once more
\begin{equation}
v=\frac{1}{2} \ln(t^2+v_0^2)-\frac{t}{v_0} \tan^{-1}\left(\frac{t}{v_0}\right)+\lambda_0t+\lambda_1,\label{a9}
\end{equation}
Hence the complete solution is
\begin{eqnarray}
e^w&=&w_0t,\nonumber\\
e^v&=&\sqrt{t^2+v_0^2}e^{\left[\lambda_0t+\lambda_1-\frac{t}{v_0}\tan^{-1}\left(\frac{t}{v_0}\right)\right]}\nonumber\\
e^u&=&u_0t\sqrt{t^2+v_0^2}e^{\left[-\left(\lambda_0-2\frac{Q_0}{\alpha_0}\right)t+\lambda_1-\frac{t}{v_0}\tan^{-1}\left(\frac{t}{v_0}\right)\right]},\label{a10}
\end{eqnarray}
with $c=\frac{Q_0}{\alpha_0}.$
Then the solution in the old variables takes the form
\begin{eqnarray}
a&=&\sqrt{t^2+v_0^2}e^{\frac{1}{2}\left[\left(\lambda_0-2\frac{Q_0}{\alpha_0}\right)t+\lambda_1-\frac{t}{v_0}\tan^{-1}\left(\frac{t}{v_0}\right)\right]}\nonumber\\
b&=&te^{-\frac{2Q_0}{\alpha_0}t}\nonumber\\
\phi&=&e^{\frac{2Q_0}{\alpha_0}t},\label{A11}
\end{eqnarray}
which is equation (\ref{k53}).

\end{document}